 \documentclass[usenatbib,useAMS,preprint,psfig2]{mn2e}
\usepackage{amsmath}
\usepackage{amssymb,amsfonts}
\usepackage{graphicx}
\usepackage{epstopdf}

\def\gapprox{\lower.4ex\hbox{$\;\buildrel >\over{\scriptstyle\sim}\;$}}
\def\lapprox{\lower.4ex\hbox{$\;\buildrel <\over{\scriptstyle\sim}\;$}}
                                                                                                 
\title{Anisotropic weak turbulence of Alfv\'en waves in collisionless
astrophysical plasmas}
                                                                                                 
\author[Luo and Melrose]
      {Qinghuan Luo and Don Melrose\\
        School of Physics, The University of Sydney, NSW 2006, Australia\\
}
                                                                                                 
\date{
          --- Received
         in original form December, 2005
        }
                                                                                                 
\pubyear{2006}
                                                                                                 
\begin{document}
                                                                                                 
\maketitle

\begin{abstract}
The evolution of Alfv\'en turbulence due to three-wave interactions is discussed 
using kinetic theory for a collisionless, thermal plasma. There are three
low-frequency modes, analogous to the three modes of compressible MHD.
When only Alfv\'en waves are considered, the known anisotropy of 
turbulence in incompressible MHD theory is reproduced. Inclusion of
a fast mode wave leads to separation of turbulence into two regimes:
small wave numbers where three-wave processes involving a fast mode
is dominant, and large wave numbers where the three Alfv\'en wave process
is dominant. Possible application of the anisotropic Alfv\'en turbulence to the 
interstellar medium and dissipation of magnetic energy in magnetars is discussed. 
\end{abstract}
                                                                                                 

\section{Introduction}

There is an extensive literature on the turbulence in the interstellar medium (ISM) 
that is based on incompressible MHD, in which case the only waves that can exist
are Alfv\'en waves. However, the ISM should be regarded as
a compressible MHD medium at frequencies below an effective collision
frequency, and as a collisionless, thermal, magnetized plasma at frequencies above
the effective collision frequency. An important qualitative change from the
incompressible case is that there are then three modes, referred to as the Alfv\'en, 
fast and slow modes in compressible MHD and as the Alfv\'en, magnetosonic and 
ion sound modes in a collisionless thermal plasma. In principle, wave-wave
interactions involving all three wave modes can contribute to the evolution of
turbulence

The conventional treatment of the evolution of turbulence in the ISM is in terms of
a cascade of Alfv\'en waves from a small wave number to a large wave number 
through three-wave interactions, as first proposed independently by~\cite{i64}
and \cite{k65}.  In the Iroshnikov-Kraichnan (IK) theory, the cascade proceeds through 
collisions of oppositely directed Alfv\'en wave packets, which leads to break-up of waves 
into smaller scales, corresponding to larger wave numbers. The IK theory predicts 
an energy spectrum $E(k)\sim k^{-3/2}$ in the incompressible, isotropic MHD approximation. 
The isotropy assumption is valid only if the large scale magnetic field in the ISM is ignored. 
Inclusion of the large-scale magnetic field changes the nature of wave-wave interactions 
and the spectrum of turbulence. An important effect is suppression of the cascade of 
Alfv\'en waves along the mean magnetic field direction due to a kinematic restriction, 
required by the frequency and wave vector matching conditions~\citep{setal83,sg94}. 
A wave cascade perpendicular to the magnetic field 
is permitted, so that the turbulence evolves in a highly anisotropic manner, 
evolving much faster perpendicular to than parallel to the magnetic field, 
as shown by numerical MHD simulation \citep{setal83}. This result was confirmed
more recently~\citep{mm95,nb96,getal00,wetal00,lg03}.
These studies lead to the conclusion that in the ideal, 
incompressible MHD approximation, three-wave interactions of Alfv\'en waves are 
allowed and lead to anisotropic turbulence with a spectrum $E(k_\perp)\sim k^{-2}_\perp$. 
However, there was some confusion in due to the three-wave matching conditions 
requiring that one mode must have a zero frequency~\citep{sg94,mm95,nb96,getal00,wetal00,lg03}.
The assumption that a wave must have a nonzero frequency, 
combined with the three-wave matching condition requiring strictly zero frequency. 
led to the conclusion that Alfv\'en turbulence cannot evolve through three-wave 
interactions~\citep{sg94}. However, this conclusion was later retracted~\citep{lg03}, 
with the zero-frequency, finite-$k_\perp$ disturbance
being regarded as an acceptable component of the (perpendicular) turbulence. 

In this paper we consider the evolutions of Alfv\'en turbulence in a collisionless, 
magnetized electron-proton plasma in the random phase approximation. 
As already noted, the MHD approximation applies in the limit where the 
wave frequency is small compared with an effective collision frequency, 
and it also requires that the wave number be small compared with the inverse mean free path. 
These conditions are not satisfied, for example, for fluctuations on sufficiently 
small length scales in the ISM~\citep{r90,sg90}. Indeed, in essentially all 
astrophysical and space plasmas of interest, waves at high wave number in 
the turbulent spectrum should be treated using kinetic theory. 
A more specific example of interest to us is the ultrarelativistic, superstrongly 
magnetized plasmas in magnetars, where magnetic energy dissipation through 
Alfv\'en turbulence is thought to be important in transient X-ray emission 
~\citep{td95,tb98}. In such a medium,
collisions play no role whatsoever, and
the properties of the Alfv\'en waves need to be treated using kinetic theory. 

An interesting formal problem arises when one applies the theory of 
wave-waves interactions in a collisionless plasma to the case where 
one of the modes has zero frequency. In the kinetic theory, the energy density 
in waves is proportional to their frequency, implying that there is no energy 
associated with the zero-frequency waves in the evolving perpendicular 
turbulence predicted by the MHD treatment. This conclusion cannot be correct, 
and the resolution of the dilemma is not obvious. Possible resolution might 
be sought by identifying the perpendicular turbulence with some linear mode,  
but we are unable to identify any relevant mode. A more favorable approach 
is to modify the three-wave matching conditions by allowing for resonance broadening. 
This approach has already been developed in the context of ion sound 
turbulence where a similar problem arises: the three-wave matching condition 
cannot be satisfied for three ion sound waves, but such three wave 
interaction is possible when a broadening in frequency due to nonlinear effects 
is taken into account~\citep{t71,t77}. The perpendicular turbulence 
is then regarded as a nonlinear quasi-mode of the medium, with the 
imaginary part of the frequency of the quasi-mode leading to 
the broadening of the resonance broadening. Any other form of nonlinear
broadening that leads to a nonzero frequency would also overcome the
problem in principle. Fortunately, in practice the zero-frequency wave
does not lead to mathematical problems because the evolution of the
turbulence does not involve its frequency explicitly.

Although we are concerned with the collisionless regime, it is convenient to
refer to the modes by their more familiar counterparts in compressible MHD.
We denote the Alfv\'en was as $A$, the magnetosonic (fast) mode as $F$, and
the ion sound (slow) mode as $S$. We are aware of no treatments that include
the $S$ mode in three wave interactions involving the $A$ and $F$ modes,
and we ignore this mode here. The simplest relevant approximation for the
remaining $A$ and $F$ modes is the cold plasma limit, and most of our
detailed results apply in this limit.
The conventional cascade involves $A\to A+A$, and we concentrate initially
on this interaction in the collisionless case. However, 
even if the initial waves were purely Alfv\'enic ($A$), three-wave interactions, 
$A\to A+F$, $A\to F+F$, would cause a mixture of modes to develop. 
As a result one also needs to consider the interactions $F\to F+F$, $F\to F+A$,
$F\to A+A$. The ratio of $F$ to $A$ in the resulting turbulent spectrum is of
interest, especially in view of the fact that the $F$ can experience significant
Landau damping, both by thermal particles and by cosmic rays.

The properties of the Alfv\'en mode and fast mode in a cold magnetized plasma are summarized 
in Sec.~2. Weak turbulence due to three-wave interactions involving three Alfv\'en waves 
is discussed in Sec.~3. The effect of including the $F$~mode is discussed in 
Sec.~4. Application to the ISM and to magnetars are discussed in Sec.~5 and~6, respectively.

\section{The Alfv\'en and fast modes}

The properties of low-frequency waves in a collisionless plasma are well known; 
the following summary of these properties is based on the treatment by~\cite{m86}. 

The general expression for the response tensor for a magnetized plasma with thermal 
distributions of electrons and ions, with thermal speeds $V_e, V_i$, involves 
modified Bessel functions with argument 
$\lambda_\alpha\equiv k_\perp V_\alpha/\Omega_\alpha$ and plasma dispersion functions 
with argument $z_\alpha\equiv \omega/\sqrt{2}k_\parallel V_\alpha$, where $\alpha=i,e$ 
denotes the species of particle with gyrofrequency $\Omega_\alpha=|q_\alpha|B/m_\alpha$, 
and where where $k_\perp,k_\parallel$ are the perpendicular and parallel wave vectors, 
respectively. The general form is approximated by assuming low frequencies, 
$\omega\ll\omega_{pi}$,  and small gyroradii, $\lambda_\alpha\ll1$. The ion sound 
speed and the Alfv\'en speed are identified as $v_s=\omega_{pi}V_e/\omega_p$ and 
$v_A=(\Omega_i/\omega_{pi})c$. The nearly incompressible limit of MHD corresponds 
to the collisionless regime $v_s\ll v_A$, which we assume to be satisfied. With 
these approximations the dielectric tensor reduces to the cold plasma approximation, 
which may be written in the form
\begin{equation}
K_{ij}=\delta_{ij}-\sum_{\alpha}{\omega^2_{p\alpha}\over\omega^2}\tau^{(\alpha)}_{ij}
(\omega),
\end{equation}
\begin{equation}
\tau^{(\alpha)}_{ij}={\omega^2\over\omega^2-\Omega^2_\alpha}
\left(\delta_{ij}-{\Omega^2_\alpha\over\omega^2}b_ib_j+i\eta_\alpha
{\Omega_\alpha\over\omega}\varepsilon_{ijl}b_l\right),
\label{eq:tij1}
\end{equation}
where $\eta_\alpha=q_\alpha/|q_\alpha|$ is the sign of the charge of 
species $\alpha$, and ${\bf b}={\bf B}/B$ is the magnetic field line direction,
assumed to be the 3-axis. 
The low-frequency approximation corresponds to $\omega\ll\Omega_i$,
where for simplicity we assume only one ionic species. The wave
properties are then found by assuming the component $|K_{33}|$ is very large,
and an expansion in $\omega/\Omega_i$ is performed. 
For present purposes, it is important to retain the corrections
involving $\omega/\Omega_i$ only in the polarization vectors. 

The dispersion relation, $\omega=\omega({\bf k})$, the polarization vector $\bf e$, 
and the ratio $R$ of the electric energy to the total wave energy for the 
Alfv\'en mode ($A$) and the magnetoacoustic (fast) mode ($F$) may be approximated by
\begin{equation}
\omega_A=|k_\parallel|v_0,
\quad 
{\bf e}_A={\bf e}_1-i
{\omega\over\Omega_i}\cot^2\theta\,{\bf e}_2,
\label{eq:dispA}
\end{equation}
\begin{equation}
\omega_F= kv_0,
\quad 
{\bf e}_F={\omega\over\Omega_i}\sec^2\theta\,{\bf e}_1
+i{\bf e}_1
\label{eq:dispF}
\end{equation}
\begin{equation}
R_A=R_F=v_0^2/2c^2,
\quad
v_0^2=v_A^2/(1+v_A^2/c^2),
\label{eq:RAF}
\end{equation}
\begin{equation}
{\bf e}_1=(\cos\phi,\sin\phi,0),
\quad
{\bf e}_2=(-\sin\phi,\cos\phi,0),
\end{equation}
with ${\bf k}=(k_\perp\cos\phi,k_\perp\sin\phi,k_\parallel)$  and
$\tan\theta=k_\perp/k_\parallel$. For $v_A^2\ll c^2$ one has $v_0^2\approx v_A^2$,
and for sufficiently low density or strong magnetic field one can have
$v_A^2\gg c^2$ one has $v_0^2\approx c^2$. Apart from the corrections
proportional to $\omega^2/\Omega_i^2$, these wave 
properties are then the same as those be 
derived using relativistic MHD theory~\citep{a83,tb98}.

Thermal corrections to the dispersion relations (\ref{eq:dispA}) and (\ref{eq:dispF}) 
arise by retaining extra terms in the expansion in 
$\lambda_\alpha\equiv k_\perp V_\alpha/\Omega_\alpha$ and 
$z_\alpha\equiv \omega/\sqrt{2}k_\parallel V_\alpha$. 
This is important when considering 
the damping of the waves. Landau damping by 
electrons dominates for the $F$-mode waves, 
giving a damping rate  $\Gamma_F\propto v_s/v_0$. 
Landau damping of the $F$ mode both by thermal electrons 
and by cosmic rays needs to be considered. 
Landau damping  for the $A$ mode is much weaker, 
$\Gamma_A\propto (\omega/\Omega_i)^2$,
and can be ignored.

\section{Weak Alfv\'en turbulence}

We apply the random phase formalism to three-wave interactions involving three $A$ waves.

\subsection{Interaction of three Alfv\'en waves}

A three-wave coupling between $A$ waves may be denoted by $A\leftrightarrow A'+A''$, 
where $A,A',A''$ denote the three waves. The resonance or matching condition, 
which express conservation of energy and momentum from a semiclassical viewpoint, are
\begin{equation}
{\bf k}={\bf k}'+{\bf k}'',
\quad
\omega({\bf k})=\omega'({\bf k}')+\omega''({\bf k}''),
\label{eq:3w}
\end{equation} 
with $\omega({\bf k})$ given by (\ref{eq:dispA}). These conditions then 
give $k_\parallel=k'_\parallel+k''_\parallel$ and $|k_\parallel|=|k'_\parallel|+|k''_\parallel|$,
which are satisfied only if one of the waves, say $A''$, has $k''_\parallel=0$, 
corresponding to zero frequency. Then $k_\parallel=k'_\parallel$ implies that three-wave 
interactions do not affect the distribution in $k_\parallel$ and affect 
only the distribution in ${\bf k}_\perp$. 

The interpretation of the zero-frequency mode has been discussed extensively in the 
MHD context~\citep{mm95,bn01,getal00,getal02,lg03}.
A specific problem arises with the zero-frequency 
mode in the kinetic theory formalism because a zero-frequency disturbance carries 
no energy, seemingly precluding any transfer of energy in the turbulent cascade. 
The resolution of this dilemma that we propose here is that the nonlinear 
wave-wave interactions produce a natural spread in wave dispersion, i.e. growth or damping due to
nonlinear interactions causes a broadening in frequency in the dispersion relations (\ref{eq:dispA}) 
and (\ref{eq:dispF})~\citep{t71,t77}. The idea is that the
frequency matching condition (\ref{eq:3w}) 
expresses energy conservation in the limit when the nonlinear transfer rate is negligible in 
comparison with the wave frequencies, and using it to infer a zero-frequency mode is inconsistent 
because wave frequencies smaller than the nonlinear transfer rate are not physically meaningful. 
An alternative version of this idea is as follows. For turbulence associated with waves in 
a specific mode, the turbulent energy is proportional to the Fourier transform of the 
correlation function of the wave amplitude, $U({\bf k},\omega)$ say. If the turbulence 
is sufficiently weak, such that the nonlinear effect on the dispersion relation is negligible, 
then one has $U({\bf k},\omega)=U({\bf k})2\pi\delta(\omega-\omega({\bf k}))$~\citep{m86}, 
where $U({\bf k})$ is the energy density per unit volume of ${\bf k}$-space. The effect of the nonlinear 
transfer rate, $\Gamma$, is to broaden the $\delta$-function distribution into a Lorentzian profile 
centred at a frequency that differs from the usual dispersion relation, $\omega({\bf k})$, by
a nonlinear frequency shift, with a width $\sim |\Gamma|$~\citep{t71}.
The inclusion of nonlinear broadening allows the three-wave matching condition to be 
satisfied by three $A$ waves without requiring one of them to have a zero frequency. 

It is convenient to introduce the frequency mismatch between the three $A$ waves:\\
\begin{equation}
\Delta\Omega=(|k_\parallel|-|k'_\parallel|-|k''_\parallel|)v_0,
\label{mismatch}
\end{equation}
For the third wave to have a nonzero frequency, due to a nonlinear
correction or due to any other effect, $\Delta\Omega$ must be nonzero.
$A$ waves that all propagate either forward or backward
cannot interact through the three-wave process. For example, if all 
three waves propagate forward $k_\parallel>0$, $k'_\parallel>0$ and $k''_\parallel>0$, 
the mismatch would be zero, $\Delta\Omega=0$, but
$\Delta\Omega\ne0$ is necessary when the nonlinear interaction is included, 
forbidding three forward-propagating waves from interacting. 
The inclusion of  $\Delta\Omega\ne0$, or $\omega''\ne|k''_\parallel|v_0$,
is important conceptually, but it turns out to be unimportant in treating 
the evolution of the $A$ waves due to three-wave interaction. 
This is because the kinetic equation for the propagating Alfv\'en 
waves does not depending explicitly on the frequency of this wave. 
One needs to appeal directly to the nonlinear broadening only when considering 
the kinetic equation for the perpendicular (zero-frequency) component of the turbulence. 
However, this is not considered in the MHD treatments, and
we do not consider it here.

\subsection{Three-wave coupling}

In a semi-classical formalism~\citep{t77,m86}, the three-wave coupling 
is described by a probability that depends on the properties of the three waves 
and on the quadratic nonlinear response of the plasma. The quadratic response 
tensor, $\alpha_{ijl}$, is written down in the Appendix. The probability can be written
as $u_{AA'A''}$ times a $\delta$-function that expresses
the matching conditions (\ref{eq:3w}), and is written below as $(2\pi)^4\delta^4(k-k'-k'')$.
A detailed calculation gives
$u_{AA'A''}=4\hbar R_AR_{A'}R_{A''}|\alpha_{_{AA'A''}}|^2/(\varepsilon_0^3\omega
\omega'\omega'')$, with $\alpha_{_{AA'A''}}=e_{Ai}e_{A'j}e^*_{A''l}\alpha_{ijl}$
the projection of the nonlinear response tensor onto the polarization vectors
of the three waves. Some further details are discussed in the Appendix.
We find
\begin{eqnarray}
u_{AA'A''}=
{2\pi r_p v^6_0\over  v^4_A}\,{\hbar\omega''\over
m_pc^2}\, \left({k_\perp c\over\omega}\right)^2
\left({\omega\over\Omega_i}\right)^4
\Biggl[2\sin\phi'\sin\phi''
\nonumber\\
-{v_0^2\over v^2_A}\left({k_\parallel\over k_\perp}\right)^2 \!\!
\bigg(\cos(\phi''-\phi')+{k_\perp\over k'_\perp}\cos\phi''\biggr)
\Biggr]^2,
\label{eq:w2}
\end{eqnarray}
where $\phi=0$ is assumed, $m_p$ is the proton mass, assuming an 
electron-proton plasma, and $r_p=r_e(m_e/m_p)$,
with $r_e=e^2/4\pi\varepsilon_0m_ec^2\approx2.6\times10^{-15}\,\rm m$ 
the classical radius of the electron. As already noted, for a strictly 
zero-frequency mode $\omega''=0$, (\ref{eq:w2}) 
implies a zero three-wave probability, but there is a corresponding factor of $1/\omega''$ that 
effectively appears when one writes down the kinetic equation for the $A$ waves, 
allowing one to use (\ref{eq:w2}) without needing to specify the value of $\omega''$. 

\subsection{Kinetic equation for Alfv\'en waves}

The evolution of the occupation number (classically, the wave action) $N({\bf k})$ of $A$-mode waves
is determined by the kinetic equation  
\begin{eqnarray}
{dN({\bf k})\over dt}&\approx&-\int{d{\bf k}'\over(2\pi)^3}\,{d{\bf k}''\over(2\pi)^3}
u_{AA'A''}({\bf k},{\bf k}',{\bf k}'')N_{A''}({\bf k}'')\nonumber\\
&&\times\Bigl[
N({\bf k})-N({\bf k}')\Bigr](2\pi)^4\delta^4(k-
k'-k'')\nonumber\\
&=& -{2\pi^2 r_pv^5_0\over v^4_A}
\left({k_\perp\over k_\parallel}\right)^2\left({\omega\over\Omega_i}\right)^4
\!\int\!{d{\bf k}''_\perp\over(2\pi)^3}
\,{U_{A''}\over m_pc^2}\nonumber\\
&&\times\Bigl[N({\bf k})-N({\bf k}')\Bigr]
\Biggl[2\sin\!\phi'\sin\!\phi''-{v^2_0\over v^2_A}
\nonumber\\
&&
\times\!\left({k_\parallel\over k_\perp}\right)^2
\!\!
\bigg(\!\cos(\phi''-\phi')+{k_\perp\over k'_\perp}\cos\phi''\!\biggr)
\Biggr]^2.
\label{eq:ke1}
\end{eqnarray}
On the left hand side of (\ref{eq:ke1}), the derivative is interpreted according 
to $d/dt=\partial/\partial t+{\bf v}_g\cdot\partial/\partial{\bf x}+
(\partial\omega/\partial{\bf x})\cdot\partial/\partial{\bf k}$, where ${\bf
v}_g=\partial\omega/\partial {\bf k}$ is the group velocity. On the right 
hand side of (\ref{eq:ke1}) the occupation number for the double-primed mode 
is $N_{A''}({\bf k}'')=U_{A''}({\bf k}'')/\hbar\omega''$, and for $\omega''\to0$ this 
term dominates over other nonlinear terms $\propto N({\bf k})N({\bf k}')$ that 
are omitted here~\citep{lg03}. Note also the mentioned cancelation 
of $\omega''$ in the product $u_{AA'A''}({\bf k},{\bf k}',{\bf k}'')N_{A''}({\bf k}'')$. 
The equality in (\ref{eq:ke1}) is derived assuming $k'_\parallel\approx k_\parallel$ and  
${\bf k}'_\perp={\bf k}_\perp-{\bf k}''_\perp$. This approximation remains 
valid provided that the natural spread in 
frequency is much smaller than the wave 
frequency. Eq~(\ref{eq:ke1}) is similar to an equation derived by~\cite{getal02} 
(cf. their Eq [10]) using the incompressible MHD formalism. 

\subsection{Natural broadening in frequency}

Nonlinear interactions lead both to a frequency shift and to wave growth or damping, giving a
nonlinear correction to the dispersion relation, and introducing a natural broadening 
in frequency, respectively. The frequency shift and growth rate can be derived from 
\begin{equation}
\delta\omega=-{R_A\over\varepsilon_0\omega}
\,{\rm Re}\,\alpha^N,
\label{eq:Dw1}
\end{equation}
\begin{equation}
\Gamma=-{R_A\over\varepsilon_0\omega}
\,{\rm Im}\,\alpha^N,
\label{eq:Gamma1}
\end{equation}
where all terms on the RHS are evaluated at the values from the dispersion relation, 
and with $\alpha^N({\bf k})=e^*_i({\bf k})e_j({\bf k})\alpha^N_{ij}$ determined by the 
quadratic response tensor~\citep{m86}. As already noted, the evolution of
$U_{A''}({\bf k}'')$ (the zero-frequency mode) requires $\Gamma\neq0$ and
the expression (\ref{eq:Gamma1}) generally leads to a nonzero growth or damping rate.

\subsection{Turbulence spectrum}

Assuming that the three-wave interaction is the dominant process in 
the turbulence, one may derive the energy spectrum in some limiting cases 
using a method similar to that used by~\cite{z84}. The kinetic equation (\ref{eq:ke1}) 
includes integrals over the azimuthal angles $\phi',\phi''$, which may be rewritten using
\begin{eqnarray}
\cos\phi'&=&{k^2_\perp+{k'_\perp}^2-{k''_\perp}^2\over2k_\perp
k'_\perp}\equiv\mu'_c,
\label{eq:cos1a}\\
\cos\phi''&=&{{k_\perp}^2-{k'_\perp}^2+{k''_\perp}^2\over2k_\perp
k''_\perp}\equiv\mu''_c.
\label{eq:cos1b}
\end{eqnarray} 
On writing $N({\bf k})\equiv N(k_\perp,k_\parallel)/k_\perp$ and 
$U_{A''}({\bf k}'')=U_{A''}(k''_\perp)/k''_\perp$, the kinetic equation reduces to
\begin{eqnarray}
&&{dN(k_\perp,k_\parallel)\over dt}
\approx-{r_p v^5_0\over \pi v^4_A}
\left({k_\perp\over k_\parallel}\right)^2\left({\omega\over\Omega_i}\right)^4
\nonumber\\
&&\times
\int dk'_\perp dk''_\perp
\,{U_{A''}(k''_\perp)\over m_pc^2}\,{{k'_\perp}\over
k_\perp k''_\perp }
 \left(1-{\mu'_c}^2\right)\left(1-{\mu''_c}^2\right)^{1/2}
 \nonumber\\
&&\qquad\qquad\times
\Biggl[N(k_\perp,k_\parallel)-{k_\perp\over k'_\perp}N(k'_\perp,k_\parallel)\Biggr],
\label{eq:ke2}
\end{eqnarray} 
where $\mu'_c$ and $\mu''_c$ are functions of $k'_\perp/k_\perp$,  $k''_\perp/k_\perp$
given by (\ref{eq:cos1a}) and (\ref{eq:cos1b}), and the terms 
involving $(k_\parallel/k_\perp)^2$
are ignored~\citep{getal02}. The integration range is limited 
by $|\cos\phi'|\leq1$ and $|\cos\phi''|\leq1$, implying 
$k'_\perp>k_\perp-k''_\perp$, $k''_\perp<k_\perp+k'_\perp$, and $k'_\perp<k_\perp+k''_\perp$.

The turbulence spectrum is derived by a standard argument based on
a dimensional analysis. Assume that the source that 
drives the turbulence and the sink where the energy is dissipated are widely 
separated in $k$.  A stationary solution with a power-law 
$N(k_\perp, k_\parallel)\sim k^{-\alpha}_\perp$ is sought from ({\ref{eq:ke2}) 
by a conformal transform~\citep{z84}. In the inertial range, where 
there is neither source nor sink, there is a constant flux of energy from smaller 
to larger $k$, and the spectrum is stationary, $d N/d t=0$. It is convenient 
to introduce dimensionless variables $\xi_1=k'_\perp/k_\perp$ 
and $\xi_2=k''_\perp/k_\perp$. Assuming 
$U_{A''}(k''_\perp)\sim {k''_\perp}^{-\beta}$, Eq (\ref{eq:ke2}) becomes
\begin{eqnarray}
{d N\over d t}&\sim&
\int d\xi_1 d\xi_2\, \xi^{-\beta-1}_2\xi_1\nonumber\\
&&\times\Bigl( 1-{\mu'_c}^2\Bigr)
\Bigl(1-{\mu''_c}^2\Bigr)^{1/2}
\Bigl(1-\xi^{-\alpha-1}_1\Bigr)\nonumber\\
&=&
-\int d\xi'_1 d\xi'_2\, {\xi'_2}^{-\beta-1}\xi'_1\Bigl(
1-{\mu'_c}^2\Bigr)
\Bigl( 1-{\mu''_c}^2\Bigr)^{1/2}
\nonumber\\
&&\times
\Bigl(1-{\xi'_1}^{-\alpha-1}\Bigr){\xi'_1}^{\alpha+\beta-4}.
\end{eqnarray} 
The second form of the integral is obtained from the first by writing 
$\xi'_1=1/\xi_1$, $\xi'_2=\xi_2/\xi_1$, $\mu'_c(\xi_1,\xi_2)
\to \mu'_c(\xi'_1,\xi'_2)$ and 
$1-{\mu''_c}^2(\xi_1,\xi_2)\to {\xi'_1}^{-2}[
1-{\mu''}^2_c(\xi'_1,\xi'_2)]$. 
Half the sum of the two forms gives 
a third form
\begin{eqnarray}
{d N\over d t}&\sim&
\int d\xi_1 d\xi_2\, {\xi_2}^{-\beta-1}\xi_1\left(
1-{\mu'_c}^2\right)\left(1-{\mu''_c}^2\right)^{1/2}
\nonumber\\
&&\times \Bigl(1-{\xi_1}^{-\alpha-1}\Bigr)\Bigl(
1-{\xi_1}^{\alpha+\beta-4}\Bigl).
\end{eqnarray}
A stationary solution exists for $\alpha+\beta=4$. For $\alpha=\beta$, corresponding to a single power-law solution, the conditions for stationarity then gives $\alpha=\beta=2$, i.e.
\begin{equation}
N(k_\perp,k_\parallel)\sim k^{-2}_\perp. 
\label{eq:spectrum1}
\end{equation}
The spectrum has the same form as for incompressible MHD~\citep{bn01,getal00,getal02,lg03}.

\section{Turbulence involving the $F$ mode}

The $F$ (magnetosonic) mode may play an important role in the Alfv\'en 
turbulence through the decay and coalescent 
processes $F\leftrightarrow A+A'$, $A\leftrightarrow F+A'$.
Such processes have no counterpart in incompressible MHD. 
Although, these processes were discussed by~\cite{lt70},
the broadening effect was not considered. In this Section we include the 
broadening explicitly and discuss the relative importance of these processes 
to three-Alfv\'en-wave interactions.

Consider the case where a $F$-mode wave interacts with two $A$ waves, 
$F\leftrightarrow A+A'$, that all propagate forward or all backward. 
It can be shown that when the  frequency spread is included the three wave 
interaction is forbidden. Let $(\omega,{\bf k})$ and $(\omega',{\bf k}')$, 
$(\omega'',{\bf k}'')$ be the wave frequency and wave vector of the $F$-mode 
wave and the two $A$-mode  waves, respectively. These satisfy the 
three-wave condition (\ref{eq:3w}). Assuming $\Delta\Omega>0$ to be a broadening 
to the three-wave frequency condition as the result of nonlinear interactions, 
for $k'_\parallel>0$ and $k''_\parallel>0$, one has
\begin{equation}
k=k'_\parallel+k''_\parallel+{\Delta\Omega\over v_0},
\quad\quad k_\parallel=k'_\parallel+k''_\parallel,
\label{eq:kkz}
\end{equation}
which gives rise to
\begin{equation}
k^2_\perp=\left({\Delta\Omega\over v_0}\right)^2
\left(1-{2\omega\over\Delta\Omega}\right).
\label{eq:kperp1}
\end{equation}
This equation is also valid when the two $A$ waves both 
propagate backward, i.e. $k'_\parallel<0$ and $k''_\parallel<0$. Since the high frequency 
waves must satisfy $\omega\gg\Delta\Omega\neq0$, the right-hand side is negative, 
forbidding $F\to A+A'$ with the two Alfv\'en 
waves both propagating either forward or backward. An analogous result applies
for compressible MHD~\citep{tb98}.
                                                                                                        
It follows that $F\to A+A'$ is allowed only when the $A$ waves are oppositely
directed, $k'_\parallel k''_\parallel<0$.  Assuming $k'_\parallel>0$ and $k''_\parallel<0$, 
in a similar way to the derivation of (\ref{eq:kperp1}), one finds
\begin{equation}
k^2_\perp\approx {4\omega'\omega''\over v^2_0}
\left(1-{\Delta\Omega\over2\omega''}\right),
\label{eq:kperp2}
\end{equation}
where $k_\perp\neq0$ for $\Delta\Omega<2\omega''$ is explicit. The general expression 
for the three-wave interaction is rather cumbersome, and we consider 
only the special case where one of the two $A$  waves has a 
low frequency, $\omega''\ll \omega$, $\omega'$. For $\phi=0$, one has
(cf. Appendix)
\begin{eqnarray}
\alpha_{FA'A''}\approx-{2\varepsilon_0e\omega^2_{pi}\over m_p}\left(
{k_\perp \omega\omega''\over\Omega^3_i}\right)\cos(\phi'+\phi'').
\label{eq:afaa}
\end{eqnarray}
The frequencies of all three waves are subject to nonlinear broadening (cf. Sect.~3.1), 
but such frequency broadening does not lead to a qualitative change in the nature 
of three-wave interactions. Although a low-frequency $\omega''$ is assumed here,  
unlike interactions of three $A$ waves considered in Sect.~3, the 
three-wave condition does not requires $\omega''\to 0$. 
The probability can be approximated by
\begin{equation}
u_{FAA'}\approx{8\pi r_pc^2v_0^4\over 
v^4_A}\,
{\hbar\omega\over m_pc^2}\,
\left({\omega''\over\Omega_i}\right)^2\,
\cos^2(\phi'+\phi''),
\label{eq:ufaa}
\end{equation}
where (\ref{eq:kperp2}) is used to eliminate $k_\perp$. In (\ref{eq:ufaa}), we assume  
$\omega''\gg\Delta\Omega$, which is relevant here as $u_{FAA'}\propto {\omega''}^2$ 
favors a moderately high $\omega''$. 
In this approximation, the frequency spread plays no role in three-wave interactions. 

An important point is that three wave interactions can convert purely 
Alfv\'enic turbulence into a mixture of $A$ and $F$
modes. The process $A\to F+A'$ creates the $F$-mode waves and the process 
$F\to A+A'$ removes them. In a steady state there is a mixture of $F$ and $A$ 
waves with the ratio determined by the ratio
of the rates of the processes $A\to F+A'$ and $A\to A'+A''$.
The process $A\to F+A'$ is described by the same probability as for
$F\to A+A'$; one simply reverses the signs of the relevant ${\bf k}$s
and uses $\omega(-{\bf k})=-\omega({\bf k})$. The probability (\ref{eq:ufaa})
continues to apply with a replacement $\phi'\to\phi=0$. 
The ratio of $F$ to $A$ waves in the turbulent spectrum can be estimated from 
the ratio of the interaction rate that generates (or destroys) the 
$F$-modes to the interaction rate for purely $A$-mode turbulence. These
rates are $\gamma_{AFA'}\propto u_{AFA'}[N_F({\bf k}')+
N_{A'}({\bf k}'')]$ and $\gamma_{AA'A"}\propto 
u_{AA'A''}N_{A''}$, respectively. Using (\ref{eq:ufaa}) 
and (\ref{eq:w2}), one finds
\begin{equation}
{\gamma_{AFA'}\over\gamma_{AA'A''}}
\sim 4\left({\omega_{pi}\over\omega}\right)^2
\left({\omega''_{FAA'}\over\omega}\right)\,\left(1+{W_F\over W_A}\right),
\label{eq:fast_condit}
\end{equation}
where $W_F$ and $W_A$ are the energy density of the $F$ and $A$ modes,
respectively. One assumes that $W_{A'}\sim W_{A}$ and $v_A\ll c$.
We distinguish between $\omega''\to\omega''_{FAA'}$ in (\ref{eq:ufaa})
from $\omega''\to\omega''_{AA'A''}$ in (\ref{eq:w2}).
We define a critical frequency, $\omega_c$, that separates two regimes:
for $\omega>\omega_c$ the rate of $A\to A'+A''$ is faster than the rate 
of $A\to F+A'$, and for $\omega<\omega_c$ the rate of $A\to F+A'$ is faster than
that of the pure Alfv\'enic process $A\to A'+A''$.
Using (\ref{eq:fast_condit}), one derives the critical frequency as
\begin{equation}
{\omega_c\over\omega_{pi}}
\sim \left({4\omega''_{FAA'}\over\omega_{pi}}\right)^{1/3}\left(1+{W_F\over W_A}\right)^{1/3}.
\end{equation}
The inner and outer scales, corresponding to frequencies $\omega_{in}$ and 
$\omega_{out}$ say, must satisfy $\omega_{in}>\omega_c>\omega_{out}$ for this 
critical frequency to be relevant. The turbulence in the regime $\omega<\omega_c$
must involve $F\to F'+F''$ as well as the above discussed processes (involving both $A$ and $F$).
Unless a dominant process is singled out, in which case 
one may use the analytical approach similar to that discussed in Sec. 3.5,
derivation of the spectrum in this regime needs a numerical 
approach and is not discussed here. 

\section{The ISM and magnetars}

In this Section we discuss possible application of the foregoing results 
to turbulence in the ISM and to magnetic energy dissipation in magnetars.
In both cases, we concentrate in the regime $\omega\ll\Omega_i$.
In the case of the ISM, this regime corresponds to a length scale well exceeding 
the ion inertial length scale $v_A/\Omega_i$ where wave damping through 
cyclotron resonance becomes important.

\subsection{Turbulence in a collisionless ISM}

The foregoing discussion leads to two qualitative features with
implications for turbulence in the ISM. One feature is that the anisotropic 
features of Alfv\'enic turbulence found in treatments based on the incompressible MHD 
also apply to the collisionless case. The other feature is that
when $F$ mode waves are included, the ratio of the rates of the processes
$F\to A+A'$ and $A\to F+A'$ to the rate of the process $A\to A'+A''$ is large
at sufficiently low frequencies and small at sufficiently high frequencies.

The presence of the $F$ mode in the turbulent spectrum is relevant to
scintillations of compact radio sources such as pulsars and 
intraday variable sources (IDVs)~\citep{r90}. Scintillations result from
density fluctuations, which are associated with the $F$ mode but not with
the $A$ mode. Hence, if there is no $F$ mode component there should be
no scintillations. Our results suggest that the $F$ mode should be present
at frequencies lower than the critical frequency 
\begin{equation}
\omega_c\sim \omega_{pi}\left({4\Delta\Omega\over\omega_{pi}}\right)^{1/3},
\label{eq:wc}
\end{equation}
where we assume $W_F\sim W_A$ and $\omega''\sim \Delta\Omega/2$.
The ISM has multiple components with a range of temperatures and
densities~\citep{mo77}. In McKee \& Ostriker's three-component
model, the ISM consists of a hot low density component (the HIM), a cold neutral
cloud component that is embedded in the former with a much smaller
filling factor, and a warm coronal component (the WIM) that surrounds the neutral cloud.
For the HIM, the typical density and temperature are $n_i\sim 3\times10^3\,{\rm m}^{-3}$
and $T\sim 5\times10^5\,\rm K$. The plasma frequency is $\omega_{pi}=73\,
(n_i/3\times10^3\,{\rm m}^{-3})^{1/2}\,{\rm s}^{-1}$.
For a typical magnetic field $B=5\times10^{-10}\,\rm T$~\citep{h87}, 
the Alfv\'en speed is $v_A=(\Omega_i/\omega_{pi})c\approx 2\times10^5\,{\rm m}\,{\rm s}^{-1}$ 
with $\Omega_i=5\times10^{-2}\,{\rm s}^{-1}$. In such a plasma, the ratio of
the thermal pressure ($n_ek_BT$) to the magnetic energy density ($W_B=B^2/2\mu_0$)
is generally less than unity,  $\beta=n_ek_BT/W_B<1$. Assuming that
the broadening is related to the ratio of the wave energy density to the magnetic energy
density through $\Delta\Omega\sim \omega(W_A/W_B)$ (e.g. Tsytovich 1977), 
one may estimate the critical frequency as
$\omega_c\sim2\omega_{pi}(W_A/W_B)^{1/2}$. There is no reliable way to estimate
the wave energy density, though there must be an upper limit  
$W_A/W_B\ll1$. For $\omega_c$ to be in the strong magnetic field regime
($\omega_c<\Omega_i$), one must have $W_A/W_B<(\Omega_i/2\omega_{pi})^2\sim 10^{-7}$ for 
the above-nominated parameters. Eq (\ref{eq:wc}) can be converted to a length scale 
$l_c=v_A/\omega_c$ for magnetic fluctuations in the parallel direction 
(relative to the mean magnetic field). The lower limit to 
$l_c$ is the ion inertial length $l_i\equiv v_A/\Omega_i$: $l_c\sim(v_A/\omega_{pi})(W_A/W_B)^{-1/2}> 
l_i\sim 4\times10^6\,\rm m$. So, the three $A$-wave process
becomes dominant only in the region near the inner scale of the turbulence
$l_i<1/k_\parallel<l_c$ and the three-wave process involving $F$ waves
dominates in the short wavelength region $1/k_\parallel\geq l_c$.

The above conclusion is valid for the WIM as well. The WIM has a relatively high density 
$n_i\sim 10^6\,{\rm m}^{-3}$, corresponding to a plasma frequency
$\omega_{pi}=1.3\times10^3\, (n_i/10^6\,{\rm m}^{-3})^{1/2}\,{\rm s}^{-1}$,
and a relatively low temperature $T\sim 5\times10^3\,\rm K$.
One then estimates $l_{c,{\rm WIM}}/l_{c,{\rm HIM}}\sim (n_{i,{\rm HIM}}/n_{i,{\rm WIM}})^{1/2}
=l_{i,{\rm WIM}}/l_{i,{\rm HIM}} \sim 0.06$, where the two components (HIM, WIM) of 
the ISM are labelled explicitly. Thus, the anisotropic Alfv\'en turbulence 
may exist in both HIM and WIM, favorably in the scale of turbulence near $l_i$.
It is emphasized here that this result is derived under the assumption of
weak turbulence and that the possibility of strong turbulence in the ISM 
remains open.

\subsection{Alfv\'en waves in magnetars}

Alfv\'en turbulence is relevant to magnetic dissipation in magnetars,
which are supercritically magnetized neutron stars (with $B\gg B_{cr}
\approx4.4\times10^{9}\,{\rm T}$}). 
Soft gamma-ray repeaters (SGRs) and AXPs are thought to be magnetars.
Bursting X-ray emission
is believed to be powered by magnetic energy not rotational energy. 
The enormous magnetic stress beneath the surface causes
cracks in the neutron star's solid crust producing elastic (shear) waves which couple to both 
the $A$ mode and $F$ mode that propagate in the magnetosphere~\citep{betal89}. 
These large amplitude waves trigger a cascade 
producing turbulence leading to plasma heating. 
The favorable region where the wave cascade occurs is the closed field line region
where perturbations at the two opposite foot points of a closed field line launch 
$A$ waves that meet head on. Since the shear waves underneath the 
surface propagate tangentially, the $A$ waves excited by the shear waves
may propagate approximately perpendicular to the field lines~\citep{betal89}.
Assuming a sharp boundary, the shear waves can couple 
to both the $A$ mode and $F$ mode. Here we only consider the former.
The typical frequency of $A$ waves can be estimated 
from $\omega_0\sim V/\Delta d=10^6(V_s/10^6\,{\rm m}\,{\rm s}^{-1})
(\Delta d/1\,{\rm m})^{-1}\,{\rm s}^{-1}$~\citep{td95}, where $V_s=(\mu/\rho)^{1/2}$ is
the shear velocity, $\mu$ is the shear modulus, $\rho$ is the crust
density, and $\Delta d$ is the length scale of the crack.
The shear velocity is typically $V_s\sim 10^6\,{\rm m}\,{\rm s}^{-1}$ for
$\mu\sim10^{29}\,{\rm J}\,{\rm m}^{-3}$ and $\rho\sim 10^{17}\,{\rm kg}\,{\rm
m}^{-3}$. In the strong magnetic field limit, one has $v_A\to c$ and the dispersion
relation $k_\parallel c\approx\omega_0$. The latter gives
$k_\parallel\sim 3\times10^{-3}\,{\rm m}^{-1}$. 

For ideal MHD to be valid one needs the electron number density
in the plasma to be sufficiently high that it can support the required
induction current that is dominant over the displacement current $\sim \omega\delta E/\mu_0c^2$,
where $\delta E\sim \delta B/c$ is the electric field of the wave. Since the maximum current it
can provide is $n_eec$, this condition can be written as
$\omega_p\gg(\omega\delta\Omega_e)^{1/2}$, which corresponds to
$n_e\gg3\times10^{24}\,{\rm m}^{-3}$ for $\omega\sim10^6\, {\rm s}^{-1}$ and 
$\delta\Omega_e=e\delta B/m_e\sim 1.7\times10^{20}\,{\rm s}^{-1}$ (for $\delta B\sim 10^{-2}B$ with
$B=10^{11}\,\rm T$). As there can be a wide range of frequencies 
due to $\Delta d$ not being well constrained, ideal MHD may not be applicable 
and the collisionless regime may be the more appropriate. 

For the wave cascade to be relevant to plasma heating in magnetars, 
the decay time of $A$ waves must be shorter than the flow time ($\sim$ 
the rotation period$/2\pi$) of the plasma in the magnetosphere.
The propagation of $A$ waves near the surface is oblique, with
\begin{equation}
{k_\perp\over k_\parallel}\approx {c\over V}\approx 3\times10^2,
\end{equation}
for $V\sim 10^6\,{\rm m}\,{\rm s}^{-1}$~\citep{betal89,tb98}.
The three-wave process involving a $F$ wave is
important at $\omega_0<\omega_c$, where the critical can be estimated
in a similar way to (\ref{eq:wc}). In the limit $v_0\to c$, one finds
\begin{equation}
\omega_c\sim {k_\parallel\over k_\perp}\left({\delta B\over B}\right)^2\Omega_i
\sim 10^{12}\left({R_0\over r}\right)^3\,{\rm s}^{-1},
\end{equation}
where $R_0=10^4\,\rm m$ is the star's radius. Here we assume 
$\Delta\Omega\sim \omega(\delta B/B)^2$, $\delta B/B=10^{-2}$ and $k_\perp/k_\parallel=300$.
The three $A$ wave process starts to dominate at $r\geq 10^2R_0$ for 
$\omega_0=10^6\,{\rm s}^{-1}>\omega_c$. In the following we consider 
the three $A$ wave process only.

Since wave decay through three $A$ wave interaction is preferentially in the 
perpendicular direction with the probability $\propto (k_\perp/k_\parallel)^2$, 
the three-wave interaction would be significantly enhanced as $k_\perp/k_\parallel$ increases. 
The decay time can be estimated from (\ref{eq:ke1}) as
\begin{equation}
t_d\sim {\beta^4_A\over8\pi^2r_pc\Delta L_\parallel}
\left({k_\parallel\over k_\perp}\right)^2\left({\Omega_i\over k_\parallel c}\right)^4
\left({W_{A''}\over m_pc^2}\right)^{-1},
\end{equation}
where $\Delta L_\parallel$ is the longitudinal (along the magnetic field) 
size of the volume concerned. Since the decay process
favours a large $k_\perp$, the decay rate increases as the turbulence becomes 
anisotropic. The typical luminosity 
of transient X-rays (regular bursts) from SGRs is about $10^{28}\,{\rm W}$.
A giant burst may reach as high as $10^{37}-10^{40}\,{\rm W}$.  
For example, the recent giant burst from SGR1806-20 on the 27th December 2004
released about $10^{40}\,{\rm J}$ in less than a second~\citep{betal04}.
Assume both high and low frequency waves are excited with the 
energy density of the latter being $W_{A''}$.
We also assume $\Delta L_\parallel\sim r$, where $r$ is the typical radial distance where
wave cascade occurs, and that $W_{A''}\sim 10^{31}\,{\rm J}/\Delta V$, where $\Delta V\sim r^3$
is the volume. For $k_\parallel/k_\perp\sim300$, and
$v_A/c\sim 7.2\times10^8(B_0/10^{11}\,{\rm T})
(n_i/10^{20}\,{\rm m}^{-3})^{-1/2}(R_0/r)^{3/2}$, where 
$R_0=10^4\,\rm m$ is the star's radius,
$B_0$ is the magnetic field on the surface and $n_i$ is the proton
number density, one finds $t_d\sim 1\,{\rm s}$ for a radius much smaller
than the radius of the light cylinder $r\sim 3.9\times10^3R_0\ll R_{LC}=cP/2\pi
=3\times10^4R_0(P/6\,{\rm s})$, where $P$ is the magnetar period.
For $W_{A''}\sim 10^{40}\,{\rm J}/\Delta V$, which may be needed for a giant 
flare, the dissipation occurs much closer to the surface, $r\sim 5\times10^2R_0$.
Wave cascade may occur much closer to the surface if 
one assumes a higher plasma density. 

One possible channel for dissipation is through Landau damping
by thermal electrons. Since the frequency of the $A''$ mode is
nonzero, one expects the wave cascade would increase $k_\parallel$ and that 
the damping becomes effective when the parallel scale becomes
sufficiently small $1/k_\parallel\sim c/\Omega_i\sim 1.8\,{\rm m}$ at a
radial distance $r\sim 3.9\times10^3R_0$ as compared to the initial
$1/k_\parallel\sim 300\,\rm m$. Efficient damping would lead to rapid plasma heating and
hence thermal X-ray emission.

\section{Discussion and conclusions}

We study three-wave interactions of $A$ waves in
the random phase formalism in the collisionless limit. We consider
weak turbulence so that energy and momentum of three interacting waves
are conserved in each interaction. It is shown that
similar to the recent result in the ideal, incompressible MHD,
three-wave interactions of $A$ waves can occur and lead to
anisotropic turbulence in the collisionless approximation.
In the incompressible formalism, three $A$ waves can interact
only when one of them has zero frequency as required by
the three-wave resonance condition. Here we treat
the zero-frequency mode as a low frequency limit
$\omega''\ll\omega\sim\omega'$ but nonzero. The three-wave resonance
condition can still be satisfied due to natural
broadening as the result of nonlinear interactions.
Since each interaction results in only a small change in 
$k_\parallel$, compared
to a change in $k_\perp$, the turbulence becomes anisotropic in the sense
that fluctuations in the wave energy density are elongated in the
perpendicular direction (in the $\bf k$ space), with spectrum
$\sim k^{-2}_\perp$. This result is similar to that derived
in the MHD formalism. Because three-wave interactions also lead
to a cascade in $k_\parallel$, dissipation of wave energy is possible through
Landau damping.

Three-wave processes involving the $F$ mode have a counterpart in
compressible MHD but not in incompressible MHD. Rather than a single
decay, $A\to A'+A''$, we also consider $A\to F+A'$ and $F\to A+A'$ which
create and destroy a $F$-mode component in the turbulence.
These processes involving the $F$ mode are most important
at the lowest frequencies, that is, near the outer scale of the turbulence.
At higher frequencies three-wave interactions involving three $A$ waves become
dominant, with the cross-over determined by (\ref{eq:wc}). 
Our result are consistent with the recent result by \cite{c05},
who considered weak compressible MHD turbulence including the fast mode under
the assumption of a constant density. 
We note that Cho \& Lazarian's (2003) recent numerical simulation of turbulence 
in a compressible MHD shows that the $F$ mode contributes negligibly to the 
energy transfer in the turbulence except near the outer scale region.
However, their calculation is in the strong turbulence regime where
the energy and momentum of three waves need not be conserved.

We discuss two applications:
Alfv\'en turbulence in the ISM, and in the magnetospheres of magnetars.
An important implication for the ISM is the anisotropic
nature of the turbulence which may be related to the observed
anisotropy in density fluctuations, as suggested by several authors~\citep{gs95,bn01}.
It is shown here that at a smaller wave number, three-wave processes
involving a $F$ wave is dominant and the process including $A$-mode waves only
is important at a large wave number. Only the $F$ mode involves density fluctuations 
that are required for scintillation. In contrast to the three $A$ wave interaction, 
there is no preferential direction (in $\bf k$) for wave cascades involving
an $F$-mode wave. The $F$ mode can be dissipated through cascade to larger wave numbers and 
damped near the inner scale region.

In the application to magnetars, the cascade time
is found to be short so that the cascade can occur well inside the
radius of the light cylinder.
Thus Alfv\'en turbulence through three-wave interactions provides
an important channel for dissipation. One possible way to dissipate 
the $A$ waves is through Landau damping as they cascade to
waves of short wavelength comparable to the inner scale of the turbulence.

The third MHD-like mode, the ion sound or $S$ mode, should be
considered in any extension of the analysis in this paper.
Although there have been some discussions of weak compressible MHD turbulence involving the
slow mode~\citep{k01}, we are aware of no detailed results that include the effect of
the ion sound mode in magnetized turbulence in the collisionless limit.
This mode is strongly (Landau) damped in a thermal plasma, and we
speculate that its inclusion might lead to an effective damping of
Alfv\'en turbulence through the process $A\to A'+S$, with the $S$
mode rapidly damped. This might lead to an efficient dissipation of magnetic
energy in the application to magnetars, but we do not discuss this
point further here.

\newpage

\appendix
\section{Three-wave probability}

In the random phase approximation, three-wave interactions $M\leftrightarrow
P+Q$, where $M$, $P$ and $Q$ represent the three modes, satisfy
the energy (frequency) and momentum (wave vector) conservation, which
is also referred to as the three-wave resonance condition, i.e.
${\bf k}={\bf k}'+{\bf k}''$ and $\omega_{_M}({\bf k})=
\omega_{_P}({\bf k}')+\omega_{_Q}({\bf k}'')$.
Using a semi-classical formalism, a collection of
waves in mode $M$ can be described by their occupation number $N_M({\bf k})$ in
the 6-dimensional phase space. Three-wave interactions are then described by
the probability 
$w_{MPQ}(k,k',k'')=u_{MPQ}(k,k',k'')(2\pi)^4\delta^4(k-k'-k'')$ with~\citep{t77,m86}
\begin{eqnarray}
u_{MPQ}(k,k',k'')&=&{4\hbar R_M({\bf k})R_P({\bf k}')R_Q({\bf k}'')\over
\varepsilon^3_0|\omega_{_M}({\bf k})\omega_{_P}({\bf 
k}')\omega_{_Q}({\bf k}'')|}
\nonumber\\
&&\times
|\alpha_{_{MPQ}}(k,k',k'')|^2
\label{eq:uMPQ1}
\end{eqnarray}
where $\alpha_{_{MPQ}}=e^*_{_Mi}e_{_Pj}e_{_Qs}\alpha_{ijs}$,
$\alpha_{ijs}$ is the quadratic response tensor, ${\bf e}_M$ is the 
polarization
of mode $M$, and $k=(\omega, {\bf k})$. The $\delta$-function contains
the three-wave matching conditions.
The quadratic response tensor in the cold plasma approximation has the
following symmetric (${\bf k}'\leftrightarrow{\bf k}''$ and
$j\leftrightarrow s$) form~\citep{m86}
\begin{eqnarray}
\alpha_{ijs}&=&-\sum_{\alpha}{q^3_\alpha n_\alpha\over m^2_\alpha}
\Biggl[
{k_r\over\omega'}\tau^{(\alpha)}_{rj}(\omega')\tau^{(\alpha)}_{is}
(\omega'')\nonumber\\
&&+
{k_r\over\omega''}\tau^{(\alpha)}_{rs}(\omega'')\tau^{(\alpha)}_{ij}
(\omega')+
{k'_{r}\over\omega}\tau^{(\alpha)}_{ir}(\omega)\tau^{(\alpha)}_{js}
(\omega'')
\nonumber\\
&&+
{k''_r\over\omega}\tau^{(\alpha)}_{ir}(\omega)\tau^{(\alpha)}_{sj}
(\omega')-
{k''_{r}\over\omega'}\tau^{(\alpha)}_{rj}(\omega')\tau^{(\alpha)}_{is}
(\omega)\nonumber\\
&&- {k'_{r}\over\omega''}\tau^{(\alpha)}_{rs}(\omega'')\tau^{(\alpha)}_{ij}
(\omega)
\Biggr],
\label{eq:aijs1}
\end{eqnarray}
with $\tau^{(\alpha)}_{ij}$ given by (\ref{eq:tij1}).  Since
$\tau^{(\alpha)}_{xz}=\tau^{(\alpha)}_{zx}=\tau^{(\alpha)}_{yz}=\tau^{(\alpha)}_{zy}=0$,
one has $\alpha_{_{MPQ}}(k,k',k'')=0$ for the Alfv\'en or $F$ modes 
propagating parallel
or antiparallel to the mean magnetic field. In this special case, they do
not interact with each other.

It can be shown that in the expansion $O(\omega/\Omega_i,\omega'/\Omega_i,
\omega''/\Omega_i)$, only $\alpha^{(4)}_{ijs}$ is important. Since $\tau_{zz}\sim O(1)$,
$\alpha^{(0)}_{ijs}$ is nonzero only for $i=j=s=z$. However, since $e_{Az}\sim
O(1/\Omega^2_\alpha)\times O(v^2_s/v^2_A)$ ($e_{Az}=0$ for a cold 
plasma), one has
$e_{Az}e_{A'z}e^*_{A''z}\alpha^{(0)}_{zzz}\sim O(1/\Omega^6_\alpha)$.
The nonzero components of $\alpha^{(1)}_{ijs}$ correspond to one or 
two of the three
indices being $z$. Then they produce terms of $O(1/\Omega^3_\alpha)\times
O(v^2_s/v^2_A)$ and $O(1/\Omega^5_\alpha)$ in $\alpha_{AA'A''}$. One 
can show that
$\alpha^{(2)}_{ijs}=0$ for $\phi=0$.
To the order $O(1/\Omega^3_\alpha)$, the relevant components are
$\alpha_{xxx}$, $\alpha_{xxy}$, $\alpha_{xyx}$, $\alpha_{yxx}$,
$\alpha_{xyy}$, $\alpha_{yxy}$, and $\alpha_{yyx}$. Since $m^2_e\Omega^3_e\gg
m^2_p\Omega^3_i$, one may
ignore the electron terms:
\begin{equation}
\sum_\alpha{q^3_\alpha n_\alpha\over m^2_\alpha}\,{1\over\Omega^3_\alpha}
\approx \sum_i{q^3_i n_i\over m^2_i}\,{1\over\Omega^3_i}+O(m_e/m_i).
\end{equation}

In the cold plasma approximation, the polarization vectors are effectively
perpendicular to $\bf B$ and then all the components of $\alpha_{ijl}$ with
any of the three indices equal to $z$ gives zero contribution to
$\alpha_{_{MPQ}}(k,k',k'')$. The remaining terms give
$\alpha^{(n)}_{ijs}=-(e^3n_i/m^2_pc\Omega^n_i)\tilde{\alpha}^{(n)}_{ijs}$
with
\begin{eqnarray}
\tilde{\alpha}^{(3)}_{xxx}\approx i
\left(k''_y\omega+2k'_y\omega''\right)c\omega,
\end{eqnarray}
\begin{eqnarray}
\tilde{\alpha}^{(3)}_{xxy}\approx{i}
\left(\omega''-\omega\right)k''_xc\omega,
\end{eqnarray}
\begin{eqnarray}
\tilde{\alpha}^{(3)}_{xyx}\approx
{i}\left(
-k'_xc\omega\omega''\right),
\end{eqnarray}
\begin{eqnarray}
\tilde{\alpha}^{(3)}_{yxx}\approx
{i}
\left(-k_xc\omega\omega''\right),
\end{eqnarray}
\begin{eqnarray}
\tilde{\alpha}^{(3)}_{xyy}\approx
\,{i}
k_yc\omega\omega'',
\end{eqnarray}
\begin{eqnarray}
\tilde{\alpha}^{(3)}_{yxy}\approx
{i\over\Omega^3_i}\,
k'_yc\omega\omega'',
\end{eqnarray}
\begin{eqnarray}
\tilde{\alpha}^{(3)}_{yyx}\approx
{i\over\Omega^3_i}\left(
\omega-\omega''\right)k''_yc\omega,
\end{eqnarray}
\begin{eqnarray}
\tilde{\alpha}^{(3)}_{yyy}\approx
{i\over\Omega^3_i}\left(
-k''_x\omega-2k'_x\omega''\right)c\omega,
\end{eqnarray}
\begin{eqnarray}
\tilde{\alpha}^{(4)}_{xxx}
\approx0,
\end{eqnarray}
\begin{eqnarray}
\tilde{\alpha}^{(4)}_{xxy}\approx0,
\end{eqnarray}
\begin{eqnarray}
\tilde{\alpha}^{(4)}_{xyx}\approx2
\left[-\omega k''_y+\omega''(k''_y-k'_y)\right]c\omega^2,
\end{eqnarray}
\begin{eqnarray}
\tilde{\alpha}^{(4)}_{xyy}\approx2
\left[\omega k''_x+\omega''(k'_x-k''_x)\right]c\omega^2.
\end{eqnarray}
We only retain terms up to $O(\omega''/\omega)$. 
The lowest order terms in the expansion give
\begin{eqnarray}
\alpha^{(4)}_{AA'A''}&=&
e^{(0)}_{Ai}e^{(0)}_{A'j}e^{(0)*}_{A''s}\alpha^{(4)}_{ijs}+
e^{(1)}_{Ai}e^{(0)}_{A'j}e^{(0)*}_{A''s}\alpha^{(3)}_{ijs}\nonumber\\
&&+
e^{(0)}_{Ai}e^{(1)}_{A'j}e^{(0)*}_{A''s}\alpha^{(3)}_{ijs}+
e^{(0)}_{Ai}e^{(0)}_{A'j}e^{(1)*}_{A''s}\alpha^{(3)}_{ijs}
\nonumber\\
&=&e^{(0)}_{A'j}e^{(0)*}_{A''s}\alpha^{(4)}_{xjs}-{i\omega\over\Omega_i}
\cot^2\theta e^{(0)}_{A'j}
e^{(0)*}_{A''s}\alpha^{(3)}_{yjs}\nonumber\\
&&+e^{(1)}_{A'j}e^{(0)*}_{A''s}\alpha^{(3)}_{xjs}+
e^{(0)}_{A'j}e^{(1)*}_{A''s}\alpha^{(3)}_{xjs},
\label{eq:aAAQ2}
\end{eqnarray}
where the polarization vector is written into the form 
${\bf e}\approx e^{(0)}+e^{(1)}$
The second equality is derived for $\phi=0$.
For $k_\parallel\sim k'_\parallel$ one has 
$|\cos\theta''|=|k''_\parallel|/\sqrt{
{k''_\parallel}^2+{k''_\perp}^2}\ll1$. Thus we have
\begin{equation}
|e^{(0)}_{A'j}e^{(1)*}_{A''s}\alpha^{(3)}_{xjs}|
\sim 
|\cot^2\theta''|={{k''_\parallel}^2\over{k''_\parallel}^2+{k''_\perp}^2}\sim 
O({\omega''}^2).
\end{equation}
The first term in Eq. (\ref{eq:aAAQ2}) is dominant, which is used to calculate
(\ref{eq:w2}). Here one assumes a single ionic component  $i=p$ (protons).

The quadratic response for $F\leftrightarrow A+A'$ can be written
as $e_{Fi}e_{Aj}e_{A's}\alpha_{ijs}\approx
e_{Fi}e_{Aj}e_{A's}\alpha^{(3)}_{ijs}$.
One may derive (\ref{eq:afaa}) using (A4)--(A11).


\end{document}